\documentclass{article}
\usepackage[a4paper,top=3cm,bottom=3.5cm,left=3.2cm,right=3.2cm]{geometry}

\usepackage[utf8]{inputenc}
\usepackage{authblk}
\usepackage{graphicx}
\usepackage{subfigure}
\usepackage{amssymb}
\usepackage{lineno}
\usepackage[english]{babel}
\usepackage{palatino, url}
\usepackage{amsmath}

\usepackage{listings}
\usepackage{graphicx}
\usepackage[rightcaption]{sidecap}
\usepackage{textcomp}
\usepackage{array} 
\usepackage{booktabs} 
\usepackage{multirow} 
\usepackage{caption} 
\usepackage{floatflt}
\usepackage{rotating} 
\usepackage{scrextend}
\usepackage{multirow}
\usepackage[numbers]{natbib}
\usepackage{soul}
\usepackage{mathtools}

\title{Threshold sensing yields optimal path formation in \textit{Physarum polycephalum}}

\author[1,2]{Daniele Proverbio}
\author[1,2]{Giulia Giordano}
\affil[1]{\small Department of Industrial Engineering, University of Trento, Trento 38123, IT }
\affil[2]{\small Email: daniele.proverbio@unitn.it; giulia.giordano@unitn.it}

\begin{document}

\maketitle

\begin{abstract}
The model organism \textit{Physarum polycephalum} is known to perform decentralised problem solving despite absence of nervous system. Experimental evidence and modelling studies have linked these abilities, and in particular maze-solving, to some sort of memory and adaptation. However, despite compelling hypotheses, it is still not clear whether the tasks are solved optimally, and which key dynamical mechanisms enable \textit{Physarum}’s impressive abilities. Here, we employ a circuital network model for the foraging behaviour of \textit{Physarum polycephalum} to prove that threshold sensing yields the emergence of unique and optimal paths that connect food sources and solve mazes. We also prove which conditions lead to alternative paths, thus elucidating how the organism achieves flexibility and adaptation in a self-organised manner. These findings are aligned with experimental evidences and provide insight into the evolution of primitive intelligence. Our results can also inspire the development of threshold-based algorithms for computing applications.
\end{abstract}
%

%\linenumbers

\section{Introduction}
Do primitive organisms follow optimal strategies? If so, which mechanisms enable their optimal behaviours? Here, we propose a circuital network model for the foraging behaviour of the acellular slime mould \textit{Physarum polycephalum}, which we employ to show that its networking and maze solving behaviour leads to optimal path formation, thus supporting previous conjectures based on empirical evidence about emergent computational behaviours in simple living systems \cite{Beekman2015,Gao2019,Adamatzky2016}. 
\textit{P. polycephalum}, a member of the \textit{Myxogastria} class of slime moulds, is a multinucleate amoeboid organism able to behave as a giant single cell and to solve problems of varying complexity to ensure survival without possessing a central neural system; thanks to these special features, \textit{Physarum} is a paradigmatic organism for the study of emergent problem-solving by brainless life forms,
and a leading specimen for investigations on motility, environmental sensing \cite{hader1984phototactic} and response to chemical and physical stimuli \cite{Ueda1975, saigusa2008amoebae}. \textit{Physarum} can be thought of as a complex network of contractile tubules, which can adapt in response to external stimuli, such as food sources or light, by adjusting the flow of internal medium through mechanochemical processes \cite{grube2016physarum, le2024physarum}. This adaptability enables \textit{Physarum} to navigate different complex environments, solve puzzles and mazes by connecting food sources via shortest paths \cite{Nakagaki2000}, link multiple food sources \cite{Nakagaki2004} while avoiding risky environments \cite{Nakagaki2007}, and optimise transport networks \cite{Jones2010,Jones2011}. Understanding how a simple organism without neural system nor central brain-led coordination \cite{Beekman2015, Reid2023} can display such problem-solving abilities would shed light on the origin of primitive intelligence and the evolution of computing in living beings. Moreover, it would provide efficient bio-inspired algorithms to solve problems, e.g., in optimization, maze exploration, or statistics \cite{jones2016applications, Adamatzky2016, zeng2025physarum}. 

Numerous models have been developed to capture various aspects of \textit{P. polycephalum}, including its life cycle, point-of-interest and foraging behaviours, nutrient relay, and maze solving \cite{oettmeier2017physarum}. Some models offer insight into oscillations and peristalsis within the cellular endoplasm (i.e., the fluid inner layer of the cytoplasm in amoeboid cells) \cite{Rodiek2015, Alim2013} that yield memory effects and directional migration \cite{Teplov2010,tachikawa2010mathematical}. Other models capture network formation using Hagen-Poiseuille law on graphs \cite{Tero2007}, reaction-diffusion \cite{adamatzky2012slime, Ghanbari2023}, cellular automata \cite{Jones2011, Wu2015, gunji2008minimal} or multi-agent approaches \cite{liu2017new, reginato2025bottom}. To various degrees, these models reproduce experimental findings, such as the reinforcement of the main veins that follow efficient paths via flows of endoplasm and nutrients \cite{oettmeier2017physarum}; they help explain how \textit{P. polycephalum} can efficiently connect food sources in different configurations, with high fault tolerance; and they reveal how simple laws and feedback mechanisms within the organism can lead to the emergence of space exploration and pruning of redundant branches.

Alternative approaches complement the insight obtained through mechanical models by modelling \textit{Physarum} as a dynamical system on a network \cite{Gao2019}, to investigate its adaptation and problem-solving ability, which requires some form of memory and ``learning" (within the context of a microbiological interpretation). The organism \textit{P. polycephalum} follows a two-step process to refine its network \cite{dussutour2024flow}: first, it extends several advancing fronts, effectively exploring the environment and comparing multiple alternative routes at the same time; then, it reacts to the local chemical gradients emanated from food and to the flow of food through its tubules, pruning the less favourable branches and only maintaining those that make the food transport network efficient. Open questions in the literature are whether this exploration-and-pruning process eventually results in an \textit{optimal} network and, if so, which key mechanisms enable optimality. It has been hypothesised \cite{siriwardana2012fast, Adamatzky2016} that optimisation is actually performed by minimising the cost of maintaining and transporting nutrients along the tubules. According to this hypothesis, problem-solving is thus the result of \textit{Physarum}'s dynamics and of an optimisation principle that minimises dissipated energy, similarly to what is observed in other systems in biology as well as in physics, such as microbiological colonies \cite{proverbio2020assessing}, self-organised mechanical systems \cite{bak1988self}, and electrical circuits \cite{Blanchini2021}.
Hence, \textit{Physarum} is an ideal organism to test the long-standing hypothesis in systems biology that problem solving can be achieved as the emergent solution to the optimisation of nonlinear dynamics, in a decentralised fashion, without requiring a specialised central nervous system.

Drawing on the analogy between gel/sol fluxes that transport nutrients and currents, \textit{Physarum}'s foraging behaviour has been modelled via a dynamical system on a network governed by fluid-dynamics laws \cite{Tero2007}. An additional step of abstraction that simplifies the system, making it analytically tractable, has been proposed in \cite{Pershin2009} and subsequently employed in \cite{Gale2015}, and relies on the analogy with circuits of bio-inspired memristors (i.e., electrical components with state- and history-dependent resistance). In fact, in its plasmodium stage, the cytoplasm of \textit{P. polycephalum} differentiates into two forms: endoplasm containing fluid sol, and viscous gel-like ectoplasm. Movement follows a process called shuttle-streaming, driven by hydrostatic pressure due to rhythmic contractions: a pressure gradient is brought about by contracting fibres in response to environmental stimuli, and the pressure potential increases up to the point at which it causes the gel to break down into sol. This drives the formation of new low-viscosity channels after a certain threshold in pressure potential is reached \cite{Rodiek2015}. Restoring initial conditions upon changes in the environment requires time to break the gel structures down again, and depends on the shape and number of the formed veins \cite{wohlfarth1979oscillatory}. These hydraulic mechanisms can be modelled as hydraulic conductivities among webs of tubules \cite{Tero2007}. Chemotaxis (i.e., movement along chemical gradients) and sensing of food sources are also known to be governed by threshold phenomena: it has been observed that the response of \textit{Physarum} to environmental stimuli is usually not gradual, but step-wise \cite{Ueda1975, Ghanbari2023}. 
Threshold responses, nonlinear dependence of sol evolution on environmental changes,  and state-dependent restoration of conditions are phenomena that bear resemblance to those occurring in memristor-based devices \cite{strukov2008missing, Gale2015}, see Fig.~\ref{fig:scheme}a. This observation has given rise to models where memristors represent the circuital counterpart of nonlinear gel/sol interactions, playing the role of memory components, while the resistance of memristors encodes the behaviour of low-viscosity channels \cite{Pershin2009}. Models based on the memristor analogy successfully capture key aspects of \textit{Physarum}'s emergent problem-solving abilities (although they are not all-encompassing \cite{Adamatzky2016, braund2016physarum}, and neglect some physiological aspects of the cellular organism \cite{latty2010food, latty2011irrational} in favour of a predominantly phenomenological representation of the shuttle-streaming process).

Experiments showing memristor-based computational behaviour by \textit{Physarum} have been successfully conducted \cite{gale2016memristive, cifarelli2015bio};
in electronics, suitable parallel circuits have been numerically shown to solve mazes \cite{Pershin2011} and experiments employing circuit-level bio-inspired approaches to solve mazes have obtained consistent success \cite{ntinas2017oscillation}. Identifying the key features that enable maze solving, and proving that the obtained solution is optimal, would cast light on the evolution of biological organisms and guide the development of bio-inspired technologies.

In this work, we propose a circuital model for \textit{P. polycephalum}, building upon the memristor network introduced in \cite{Pershin2011,Pershin2009}, and employ it to prove that threshold effects in shuttle-streaming yield minimum-path formation in foraging and maze-solving. Nonlinear responses to chemical and pressure gradients enable optimal network formation, irrespective of the configuration of the considered environment. We also show that the circuital dynamic model explains and qualitatively reproduces the final \textit{Physarum} configurations observed empirically, where one or more branches link food sources, after a transient period during which several probing branches are initially generated and subsequently dismissed, except for those directly connecting the food sources. \textit{Memory}, \textit{learning} and \textit{problem solving} are thus global properties that, as we formally prove, can emerge from nonlinear local decisions over a complex network.

\section{The Model: Circuit Analogy and Asymptotically Stable Steady State}
\label{sec:2}

\begin{figure}[t]%[tbhp]
\centering
\includegraphics[width=0.95\linewidth]{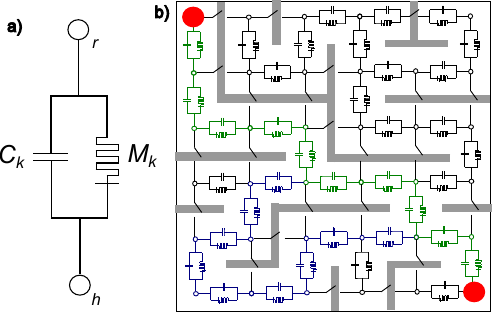}
\caption{\small Circuital model of \textit{P. polycephalum} to map a maze into a circuit. a) Schematic representation of the circuit component $k$, with capacitance $C_k$ and memristive element $M_k$, connecting nodes $h$ and $r$ of the network. b) Example of maze with walls (grey) and with a source and a sink (red circles) at its entry and exit, respectively representing \textit{Physarum} and a food source to be reached.
The maze is mapped into a network of circuit components, where white circles are associated with nodes and circuit components with links. Encoding the maze topology means having open switches (hence, disconnection) in correspondence to maze walls. In green, the shortest path connecting source and sink; in blue, the additional portion of an alternative (longer) path. }
\label{fig:scheme}
\end{figure}

When it enters a maze that contains food sources, \textit{Physarum polycephalum} tends to first explore most of the environment randomly until it encompasses and connects all food sources, and then refine its branches thanks to the feedback provided by the backward flow of nutrients \cite{Nakagaki2000, Nakagaki2004}.
To describe \textit{Physarum}'s process of exploration and refinement on a maze, the dynamical model in \cite{Tero2007} was the first to associate the maze with a graph, where the turns of the maze correspond to the graph's nodes and are connected by the graph's links; the model describes the network formation process of \textit{Physarum} through the dynamics of fluxes of nutrients along a network of tubules, each associated with a link in the graph. Sources and sinks are assumed to be present in the maze, usually corresponding, respectively, to the main protruding body of \textit{Physarum} and to the food source placed at an exit of the maze, and thus set the boundary conditions for the fluxes of nutrients. At each turn of the maze, \textit{Physarum} refines its path by responding to environmental stimuli and pressure gradients: the model in \cite{Tero2007} relies on a hydraulic interpretation of the network dynamics, where the links are akin to tubules with a certain radius and the flux over a link is governed by a viscosity- and radius-dependent conductivity that evolves over time depending on the flux on the link.
The flux-dependent hydraulic conductivity in \cite{Tero2007} plays the role of a memory component whose nonlinear and time-varying behaviour is similar to that of memristors \cite{yang2008memristive,strukov2008missing}.

The memristor-based analogy leads to circuital models of \textit{Physarum} where electrical currents (akin to fluxes of nutrients) flow through circuital components (akin to the tubules). In the simplest model of \textit{Physarum} based on a network of memristors \cite{Pershin2011}, the considered maze is mapped into a grid of circuit components, each formed by a memristor in series with a switch (approximated by a field-effect transistor). Switches encode the maze topology: closed switches connect memristors so that \textit{Physarum}'s tubule continues, while open switches correspond to disconnections, e.g., due to walls (see Fig.~\ref{fig:scheme}b).  To make the circuital component more biologically plausible, Pershin et al. \cite{Pershin2009} proposed to add a capacitor, which can store a ``charge" of endoplasm, in parallel to the memristor; see also \cite{ntinas2017modeling} for subsequent applications of the idea. Together, the capacitor and the memristor in parallel encode the response of \textit{Physarum}'s tubules to fluxes (interpreted as currents) of nutrients, subject to pressure differences (interpreted as electric potential differences between the capacitor plates), modulated by nonlinear responses (modelled by the memristor).

This circuital analogy has several consequences for the understanding and application of \textit{Physarum}'s dynamics. On the one hand, it condenses morphological characteristics of tubules into the properties of a circuital component \cite{Pershin2009,gale2016memristive, Adamatzky2016}, losing precise information about the size and shape of \textit{Physarum} (while information about the network topology is encoded in a graph, as discussed later). In addition, differently from mechanochemical models \cite{Ghanbari2023}, models based on the circuital analogy are phenomenological in nature, and hence their parameters do not exactly correspond to measurable biological quantities and cannot be directly associated with mechanical and biological properties of \textit{Physarum}. On the other hand, models of \textit{Physarum} as a memristive circuit have several advantages. First, they retain the basic properties of topology and dynamics, enabling to focus on the dynamical mechanisms that underlie the construction of transport networks and on assessing their optimality. Second, they are analytically tractable and can thus be employed to formally prove the emergence of optimality driven by local cost minimisation. Third, their underlying analogy contributes to the powerful connection between systems biology and electronics, which over decades has enabled improved understanding of the evolution and dynamics of biological organisms, beyond reductionist approaches \cite{friboulet2005systems}, as well as bio-inspired insights and applications, including \textit{Physarum}-inspired circuits for problem solving and neural-like operations \cite{Gale2015, braund2016physarum, ntinas2018coupled}.

\begin{figure}[t]%[tbhp]
\centering
\includegraphics[width=.9\linewidth]{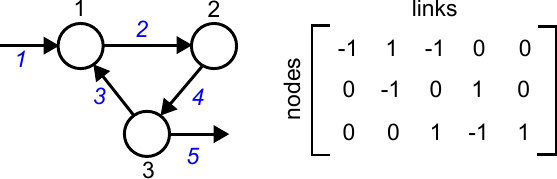}
\caption{\small A network graph (with nodes labelled in black and links labelled in blue) and its corresponding incidence matrix.}
\label{fig:incidence_matrix}
\end{figure}

We thus consider the memristor-based circuital modelling framework \cite{Pershin2009,Pershin2011} to capture the network dynamics of \textit{P. polycephalum}. Given a maze that contains food sources, we assume that \textit{Physarum} performs its initial exploration to cover it, and then starts to transport nutrients and to refine its network; as shown in \cite{Pershin2011, Tero2007}, this corresponds to mapping the maze topology into a network of circuit components, each corresponding to a segment of \textit{Physarum} (Fig.~\ref{fig:scheme}b).
Following \cite{Pershin2011}, we consider a network where nodes and links are placed so as to give shape to the maze turns, and nodes include among them the food sources to be connected.
% \cite{liu2017new}
Obstacles such as walls and barriers correspond to disconnections among nodes; instead of using field-effect transistors, as in \cite{Pershin2011}, or LR components, as in \cite{Pershin2009}, to approximate switches, we directly include disconnections (open switches) in the considered graph (see, e.g., Fig.~\ref{fig:scheme}b); hence, the basic circuit component is the one shown in Fig.~\ref{fig:scheme}a.
Mathematically, \textit{Physarum}'s network topology can be described by a graph $\mathcal{G}=(\mathcal{N},\mathcal{L})$, where the links in set $\mathcal{L}=\{1,\dots,m\}$ model memristive components that represent \textit{Physarum} branches, and
the nodes in set $\mathcal{N}=\{1,\dots,n\}$ represent where branches join (as in Fig.~\ref{fig:scheme}).
Node $1$ corresponds to the entry of the maze,
where \textit{Physarum}'s main body begins its elongation and the corresponding contractile activity, and node $n$ to the exit of the maze, where the food source to be reached is located.
The generalised node-link incidence matrix for graph $\mathcal{G}$ is given by $B \in \{-1, 0, 1 \}^{n \times m}$, whose $n$ rows are associated with nodes and whose $m$ columns are associated with links.
To compute $B$, for each $k$-th column, we assign an arbitrary direction to link $k=(h,r)$, so that the starting node is the $h$-th and the ending node is the $r$-th; then, $B_{hk}=1$ and $B_{rk}=-1$, while all other entries in the column are zero. Links coming from the external environment into the entry of the maze
(i.e., associated with \textit{Physarum} entering the maze)
have a single nonzero entry (equal to $-1$) corresponding to their ending node, and links leaving the exit of the maze have a single nonzero entry (equal to $1$) corresponding to their starting node. An example is provided in Fig.~\ref{fig:incidence_matrix}. We assume that the graph is connected, both internally and with the external environment; this assumption is reasonable as \textit{Physarum} is a unique multinucleate organism, and its syncytium remains connected over the network formation process \cite{oettmeier2017physarum}. Consequently, $B$ has full row rank. As long as this assumption is satisfied, our results hold independent of the maze topology encoded in $B$: this allows us to generalise our results to any maze, without resorting to experiments (\textit{in vitro} or \textit{in silico}) across all possible configurations. The maze topology, after the initial exploration by \textit{Physarum}, is thus mapped into a network of circuit components \cite{Pershin2011}. For homogeneous environments, without mazes or obstacles, we can consider a grid topology as in
\cite{ntinas2017oscillation},
which reflects cytosol's motile units \cite{Ghanbari2023}. Note that the maze topology is captured by the graph's incidence matrix $B$, while the flow of ectoplasm in \textit{Physarum} is encoded by the electrical currents in the circuit analogy.

\begin{figure}[t]%[tbhp]
\centering
\includegraphics[width=.75\linewidth]{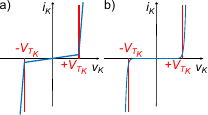}
\caption{\small In red, the ideal threshold function with threshold potential $V_{T_k} = 9$. In blue, two examples of threshold-like characteristic functions $i_k = M_k(v_k)$: a) the piecewise linear function in \eqref{eq:char_func1} with $\alpha=0.05$ and $\beta=5$; b) the smooth function in \eqref{eq:char_func2} with $j=10$.}
\label{fig:characteristic_func}
\end{figure}

We model the flow dynamics building upon the circuital representation introduced in \cite{Pershin2011,Pershin2009} and discussed above, which neglects some physiological characteristics of the biological organism to focus on endoplasm ``currents" that vary over time depending on ``potential differences" (representing pressure gradients).
We denote by $i_{M_k}$ the current of endoplasm flow \cite{Pershin2011} through the memristor on the $k$-th link of the graph, connecting nodes $h$ and $r$, and by $i_{C_k} = \frac{d}{dt}\left[C_k (v_h - v_r) \right]$ the current through the capacitor that stores the endoplasm inside a vein on the $k$-th link, in parallel to the memristor, as in \cite{Pershin2009}.
Kirchhoff's law yields a total current $i_k=i_{M_k} + i_{C_k}$ on the $k$-th link. 
The memristor can be seen as a nonlinear resistor \cite{martinsen2023bioimpedance,yang2008memristive}, so that $i_{M_k} = M_k (v_h - v_r)$, where $M_k$ is the inverse of an on/off resistance \cite{yang2008memristive}.
Recent experimental evidence has observed nonlinear current-voltage responses in \textit{Physarum} \cite[Fig.~2]{schmidt2025electrical}, which supports the choice of modelling memristors in our network as nonlinear resistive components, because it suggests that the key properties of \textit{Physarum} are rooted in its nonlinear response to the applied voltage, more than on memristive behaviours associated with memory and time dependency.
Then,
\begin{equation}
    i_k = M_k (v_h - v_r) + \frac{d}{dt}\left[C_k (v_h - v_r) \right],
    \label{eq:dyn_eq_1}
\end{equation}
where $v_h - v_r$ is the difference in stimuli applied to \textit{Physarum}'s branch (a potential difference, in the circuital analogy, akin to a pressure gradient in the hydraulic representation \cite{Tero2007}), $C_k$ is an intrinsic capacitance (related to the capacity of a vein to store gel/sol units) and $M_k \colon \mathbb{R} \to \mathbb{R}$ is a characteristic function that encodes the physical/chemical memory mechanism, e.g., gel breaking down into sol after hitting thresholds of pressure gradients \cite{matsumoto2008locomotive}, or the nonlinear interplay between the mechanics of the cytoplasm and 
flow velocity \cite{Ghanbari2023}. In the circuital representation, $M_k$ corresponds to the inverse of the memristive resistance, which describes the change in the memristor state \cite{yang2008memristive}; we assume it is possibly nonlinear, odd, locally Lipschitz and monotonically increasing
(or, more in general, non-decreasing). Ideally, $M_k$ is the inverse of a threshold step function
\begin{equation}
    \theta_k(i_k) = \begin{cases}
                V_{T_k},     &   \mbox{for } i_k > 0, \\
                - V_{T_k},   &   \mbox{for } i_k < 0, \\
                \text{any } v_k \in [-V_{T_k}, V_{T_k}], &  \mbox{for } i_k = 0,
                \end{cases}
                \label{eq:M_k}
\end{equation}
where $V_{T_k}$ is a (possibly link-specific) threshold potential; hence, it is $M_k(v_k)=0$ for $|v_k| \leq V_{T_k}$, $M_k(v_k)=+\infty$ for $v_k > V_{T_k}$ and $M_k(v_k)=-\infty$ for $v_k < -V_{T_k}$ (red in Fig.~\ref{fig:characteristic_func}). Since the ideal characteristic function is not physically feasible, it can be approximated by piecewise linear functions
\begin{equation}
    M_k(v_k) = \beta v_k - 0.5 (\beta - \alpha)(|v_k + V_{T_k}| - |v_k - V_{T_k}|),
    \label{eq:char_func1}
\end{equation}
where $\alpha$ and $\beta$ are positive constants.
Eq.~\eqref{eq:char_func1} is the negative of the function considered in \cite[Fig.~2b]{Pershin2009}.
The ideal characteristic function can also be approximated by smooth functions such as
 \begin{equation}
     M_k(v_k) = (v_k / V_{T_k})^{2j+1},
     \label{eq:char_func2}
 \end{equation} 
 with a large enough constant $j \in \mathbb{N}$ \cite{Blanchini2021}. Both examples are shown, in blue, in Fig.~\ref{fig:characteristic_func}. As long as our qualitative assumptions on $M_k$ are satisfied, the results we are going to prove hold independent of the exact functional form and of the numerical values of the coefficients.

The terminal potentials $v_k$, associated with the nodes, can be stacked in vector $v = [v_1 \dots v_n]^\top \in \mathbb{R}^n$. Then, using the previously defined incidence matrix $B$ and denoting by $B_k$ the $k$-th column of $B$, we can rewrite \eqref{eq:dyn_eq_1} as $i_k = M_k \left(B_k^\top v \right) + C_k B_k^\top \dot{v}$, since $C_k$ is assumed to be constant.
Also the flows $i_k$ at the links can be stacked in vector $i = [i_1 \dots i_{m}]^\top \in \mathbb{R}^m$, while capacities and characteristic functions can be stacked in matrix $C = \text{diag} \{C_1, \dots, C_{m} \} \in \mathbb{R}^{m \times m}$ and in vector $M(\cdot) = [M_1(\cdot) \dots M_m(\cdot)]^\top \in \mathbb{R}^m$ respectively. We can then express all the flows as
\begin{equation}
    i = M \left(B^\top v \right) + C B^\top \dot{v}.
    \label{eq:dyn_mod2}
\end{equation}

Moreover, at each node $h$, we have the flow conservation condition $B^hi - d_h = 0$ (endoplasm is not created or destroyed), where $B^h$ denotes the $h$-th row of $B$ and $d_h$ is the $h$-th entry of the input flow vector $\bar{d} = [d, 0, \dots, 0]^\top \in \mathbb{R}^n$ that corresponds to \textit{Physarum} entering the environment. Overall,
\begin{equation}
    B i - \bar d = 0.
    \label{eq:curr_bal}
\end{equation}
Taken together, Eqs.~\eqref{eq:dyn_mod2} and~\eqref{eq:curr_bal} describe the complete dynamics of \textit{Physarum}'s network, in terms of flows of endoplasm and chemical or pressure potentials. Substituting $i$ from \eqref{eq:dyn_mod2} into \eqref{eq:curr_bal} and rearranging, given that the square matrix $BCB^\top$ is non-singular because $B$ is full rank, yields the complete dynamical model
\begin{equation}
    \dot{v}(t) = - \left[B C B^\top \right]^{-1} \left[B M \left(B^\top v(t) \right) - \bar{d} \right].
    \label{eq:model_final}
\end{equation}

The considered mathematical framework, which relies on the circuital analogy between \textit{Physarum}'s foraging dynamics and the dynamics of electrical networks of memristive components, leads to a mathematical model \eqref{eq:model_final} where the differential equation has the same structure as that of models adopted in \cite{Blanchini2021} to describe the flow of current between sources and sinks (including lightning discharge) in electrical networks. The fact that different physical entities (a biological organism and an electrical circuit) can be described by dynamical equations with the same mathematical form, although the interpretation and the biophysical meaning are completely different, is a fascinating instance of the universality of mathematics as a language to describe phenomena, and allows us to leverage the results derived in \cite{Blanchini2021} for lightning discharge to better understand foraging behaviour of multinucleate acellular organisms in biology.

We now show that system \eqref{eq:model_final} asymptotically converges to the stable steady state $\bar{v}$, such that
\begin{equation}\label{eq:equilibrium}
\left[B C B^\top \right]^{-1} \left[B M \left(B^\top \bar v \right) - \bar{d} \right]=0.
\end{equation}
Following \cite{Blanchini2021}, we consider $x(t) = v(t) - \bar{v}$ and substitute for $\bar{d}$ using the equilibrium condition in \eqref{eq:equilibrium} to obtain
\begin{equation}
    \dot{x}(t) = -\left[B C B^\top \right]^{-1} B \left[M \left( B^\top(x(t) + \bar{v}) \right) - M \left( B^\top \bar{v} \right)  \right].
    \label{eq:deviations}
\end{equation}
The asymptotic stability of the steady state is guaranteed by the positive definite Lyapunov function $U(x) = \frac{1}{2} x^\top B C B^\top x$ (the energy stored in the capacitors in the circuital analogy, proportional to the stored gel/sol units inside a vein of \textit{Physarum}), whose Lyapunov derivative $\dot{U}(x) = x^\top B C B^\top \dot{x}$ is negative definite. In fact, since $M$ in \eqref{eq:deviations} is a vector of increasing functions, we can write 
$M \left( B^\top(x + \bar{v}) \right) - M \left( B^\top \bar{v} \right) = \Delta \left(v(x) \right)B^\top x$,
where the diagonal matrix $\Delta(v)$ has positive continuous functions on its diagonal \cite{blanchini2016compartmental, blanchini2019network,Blanchini2021}. We can thus rewrite \eqref{eq:deviations} as
$\dot{x}(t) = - \left[B C B^\top \right]^{-1} B \Delta (v(x(t))) B^\top x (t)$;
then, substituting $\dot x$ in the expression of the Lyapunov derivative yields $\dot{U}(x) = - x^\top B \Delta(v) B^\top x < 0$ for all $x \neq 0$, which ensures asymptotic stability of the steady-state vector $\bar v$.

The fact that the considered circuit of memristors converges to an asymptotically stable configuration supports its suitability to describe the foraging behaviour of \textit{Physarum}, which evolves towards a steady-state transport network as observed in experiments \cite{Tero2007, adamatzky2012slime, reid2013solving}.

\section{Numerical Examples}
\label{sec:numerics}

\begin{figure}[t]%[tbhp]
\centering
\includegraphics[width=0.957\linewidth]{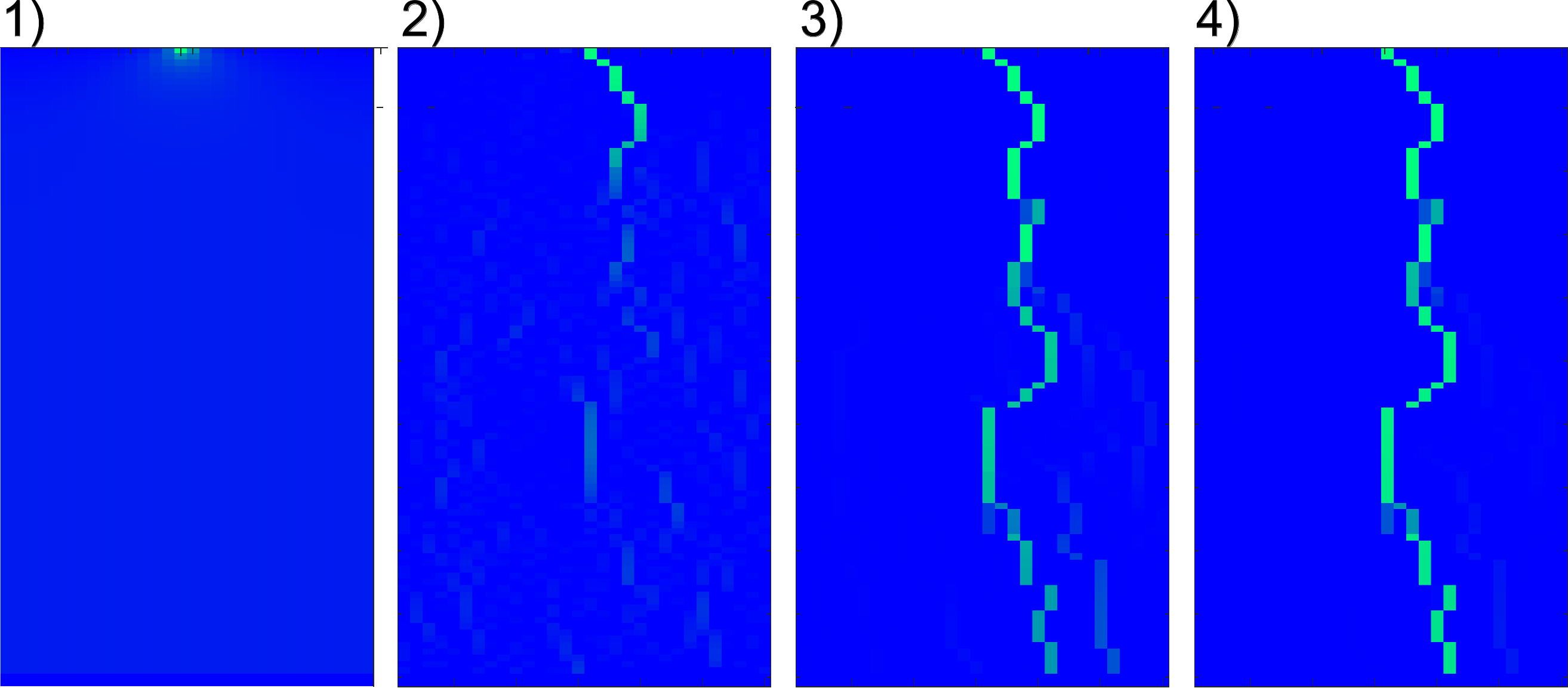}
\caption{\small Example of time evolution in scenario (b). Colour encodes the relative abundance of cellular gel, as per model \eqref{eq:model_final}, and thus highlights \textit{Physarum}'s path from top to bottom. 1) \textit{Physarum} starts spreading from its entry point ($t=100$), 2) extends ($t=1500$) and 3) probes the environment with several branches ($t=2900$) that are eventually pruned, until 4) only those (ideally, one single branch) most efficiently connecting the entry point  with the food sources remain ($t=4300$).}
\label{fig:sims_transient}
\end{figure}

\begin{figure}[t]%[tbhp]
\centering
\includegraphics[width=1\linewidth]{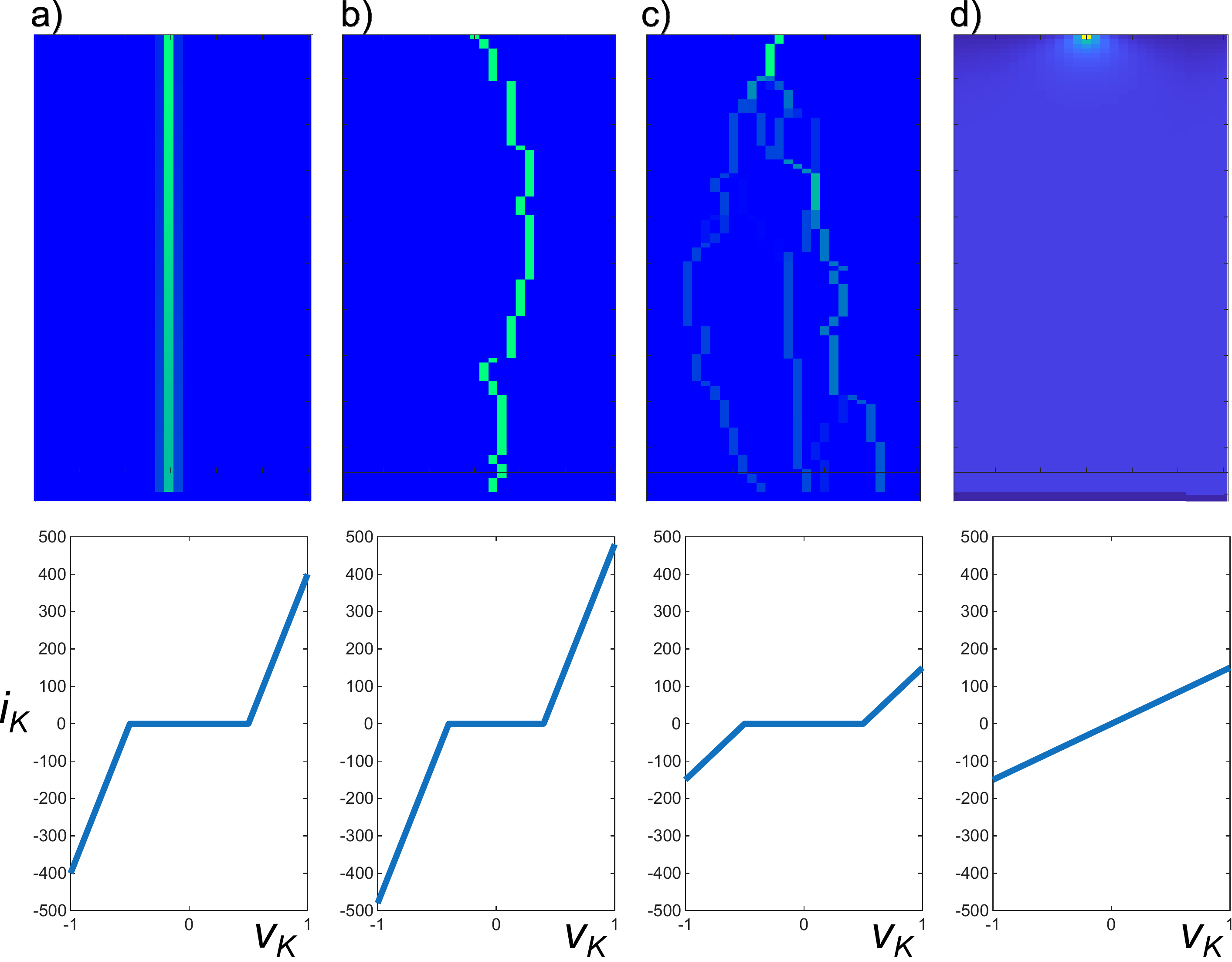}
\caption{\small Top row: Steady-state solutions for the four scenarios (a)-(d) described in Sec.~\ref{sec:numerics} in the respective panels a)-d). Colour encodes the relative abundance of cellular gel, as per model \eqref{eq:model_final}, and thus highlights \textit{Physarum}'s path from top to bottom. Note, in a) and b), the formation of a single path; in c), the formation of multiple branches; in d), the homogeneous diffusion in the whole environment. Bottom row: characteristic functions used for the simulations; a)-c) are functions of the form \eqref{eq:char_func1}, progressively more different from an ideal threshold function; d) is a linear function.}
\label{fig:sims}
\end{figure}

Simulating the memristor circuital model allows one to observe various steady-state configurations, close to empirical  \textit{Physarum}'s configurations \cite{ntinas2017modeling, Pershin2011}, which are elicited by different parameter settings. To look at several example behaviours for the model, and build an intuition of what can happen as \textit{Physarum} performs shuttle-streaming, Fig.~\ref{fig:sims_transient} shows an example of evolution towards the steady-state, while Fig.~\ref{fig:sims} shows several steady-state configurations that will be analysed later. To generate the figures, we have simulated system \eqref{eq:model_final} with an ODE solver with $C_k= 1$ for all $k$, taking $\bar{d}$ as a vector with a single non-zero entry to represent \textit{Physarum} entering the environment, using a $100 \times 30$ grid network with a 4-neighbours topology to discretise the 2D environment and construct matrix $B$ as explained above, and placing a food source spreading over the whole bottom of the environment (so that $B$ has columns with a single non-zero entry equal to 1 corresponding to the bottom nodes).
For $M_k$, we consider the piecewise expression in \eqref{eq:char_func1}, with $\alpha = 10^{-5}$.

We investigate different scenarios. (a) A homogeneous environment with $V_{T_k} = 0.5$ for all $k$ and $M_k$ close to the ideal threshold function, with $\beta = 800$. (b) A heterogeneous environment, where heterogeneities are caused by randomly scattered obstacles to the activation of chemotaxis or shuttle-streaming (cf., \textit{e.g.}, \cite{proverbio2024chemotaxis}); hence, at each $k$-th link we randomly select $V_{T_k} \in [0.5 - \delta; 0,5 + \delta]$ from a uniform distribution. We set $\delta = 0.4$. Also, we keep $\beta = 800$ so that $M_k$ is close to the ideal threshold function. (c) A heterogeneous environment set up as in (b) and $M_k$ deviating from the ideal threshold function, with $\beta=300$. (d) A homogeneous environment where $M_k$ is \textit{not} a threshold function, but a linear one: $M_k(v_k) = a\cdot v_k$ for all $k$, where $a=150$ is a scaling factor to get a magnitude of $i_k$ comparable to the other cases.
We simulate all scenarios until convergence to the steady state, \textit{i.e.}, the condition where $v(t)$ (which represents gel/sol potentials in space, thus identifying where the mould is) does not change any longer. We run the solver with time step $\Delta t=1.26 \cdot 10^{-3}$ up to $T=7936$ time units.

Fig.~\ref{fig:sims_transient} shows an example of time evolution for a run of scenario (b): \textit{Physarum} starts at the entry point, elongates and probes the environment with multiple branches, and eventually settles on a steady-state single path connecting the entry with the food sources; the final path is reached already at $t=4300 < 7936 = T$. Convergence to a steady state, as predicted by the result in Sec.~\ref{sec:2}, can be observed in all the considered scenarios, and the corresponding steady-state solutions are reported in Fig.~\ref{fig:sims}, along with the corresponding functions $M_k$ used in the simulations.
In scenario (a), as expected, \textit{Physarum} asymptotically follows a straight line (Fig.~\ref{fig:sims}a), which is the minimum path in a homogeneous environment. Fig.~\ref{fig:sims}b shows the steady-state solution of a run in scenario (b) that differs from the one in Fig.~\ref{fig:sims_transient}, because the randomly selected thresholds $V_{T_k}$ are different in the two cases and hence lead to two different steady-state paths: in both Fig.~\ref{fig:sims_transient} (panel 4) and Fig.~\ref{fig:sims}b, \textit{Physarum} avoids high-resistance areas and effectively connects to the food sources with a unique path.
Conversely, in scenario (c), when the characteristic functions significantly differ from ideal thresholds, the steady-state network is formed by several branches, as shown in Fig.~\ref{fig:sims}c. Finally, in scenario (d), when the characteristic functions are linear, \textit{Physarum} diffuses homogeneously over the whole environment (Fig.~\ref{fig:sims}d, note the shades in the colour coding). 

The performed computational analysis suggests the hypothesis that \textit{Physarum} self-organises along optimal paths connecting to food sources when the response function is close to an ideal threshold function; and the closer it is, the more likely the formation of a single path is. In fact, choosing  $M_k$ as linear functions in scenario (d) yields homogeneous diffusion; choices of $M_k$ that deviate from the ideal threshold function, as in scenario (c), lead to the formation of coexisting alternative branches, while close-to-ideal threshold functions, in scenarios (a) and (b), lead to the formation of a single path.
We also hypothesise that the obtained steady-state path is optimal, as it minimises a cost that takes into account the threshold potential needed to initiate the flow of endoplasm through a link. In the next section, we verify these hypotheses through formal mathematical analysis, in the considered modelling framework.

\section{Optimal Steady-State Path}\label{sec:opt_path}

We employ the modelling framework relying on the circuital analogy discussed in Sec.~\ref{sec:2} to:
\begin{itemize}
\item formally prove that the steady-state path chosen by \textit{Physarum polycephalum} to connect to food sources can be seen as the emergent solution to an optimisation problem and, in particular, as the \textit{minimum} path, or \textit{optimal} path, that minimises a physiological cost functional;
\item identify the exact form of the cost functional to be minimised so as to reproduce \textit{Physarum}'s foraging behaviour;
\item show that the presence of a threshold response is a necessary dynamical mechanism for the formation of optimal paths.
\end{itemize}
To derive analytical results and mathematical proofs,  we leverage the observation that the circuital analogy for \textit{Physarum} yields a mathematical model \eqref{eq:model_final} with the same structure as the electrical network model studied in \cite{Blanchini2021} and thus employ the same formalism.
We first show the optimality of the steady-state path; in Sec.~\ref{sec:linear}, we prove the necessity of nonlinear characteristic functions to achieve minimum path formation, while in Sec.~\ref{sec:branches} we consider the effect of non-ideal threshold responses.

In view of our assumptions, each function $M_k$ is invertible, and admits an inverse $g_k \doteq M^{-1}_k$ that is monotonically increasing. Hence, the function
 \begin{equation}
     f_k \colon y \mapsto \int_0^y g_k(s)ds
     \label{eq:condition}
 \end{equation}
is well defined in $(-\infty,\infty)$ and is continuously differentiable; also, $f_k$ is strictly convex if $M_k$ is increasing (respectively, convex if $M_k$ is non-decreasing), and hence $f_k'' = g_k' > 0$ (respectively, $f_k'' = g_k' \geq 0$)
almost everywhere.
Then, we can use the same intuition as in \cite{Blanchini2021} to introduce the cost functional
\begin{equation}
    J(i) \doteq \sum_{k=1}^m f_k(i_k) \doteq \sum_{k=1}^m \int_0^{i_k} g_k(I) dI.
    \label{eq:functional}
\end{equation}
$J(i)$ has the dimension of a power, since $g_k=M_k^{-1}$ is a difference of potentials (pressure potentials in \textit{Physarum}, electrical potentials in the circuit analogy) and $dI$ has the dimension of a current (current of flowing endoplasm in \textit{Physarum}, electrical current in the circuit analogy). In practice, $J(i)$ quantifies the energetic cost of maintaining a tubule (or vein) of \textit{Physarum}, as it represents the power (quantity of energy per unit time) that is required to push the flow of endoplasm through all links across a certain path.
Next, we show that the endoplasm flows in the network, sustained by a constant $\bar{d}$ and at the equilibrium ($\dot{v} = 0$), solve the optimisation problem
\begin{align}
\begin{split}
    \min \quad & \sum_{k=1}^m f_k(i_k)\\
            \text{subject to } \quad & Bi = \bar{d},
\end{split}
\label{eq:optimization}
\end{align}
that is, that the steady-state path is the one that minimises the effort (in terms of consumed energy per unit time) needed to maintain or create a branch. This result offers a rigorous mathematical proof to the long-standing conjecture in the literature that \textit{Physarum} solves mazes along \textit{shortest} paths \cite{Tero2007,siriwardana2012fast,Beekman2015}. In particular, our result shows that the path selected by \textit{Physarum} is the \textit{least energetically costly}: if the energy needed to stream gel flows through a branch is only proportional to the length of the branch itself, the optimisation problem yields the geometrically shortest path; if the required energy also depends on other stimuli, such as light, that may change the mechanochemical properties of the gel/sol medium and thus the power needed to build and sustain it \cite{grube2016physarum, le2024physarum}, then the \textit{shortest} path is indeed the \textit{least energetically costly} according to the cost functional in \eqref{eq:optimization}.

Let us now consider the solution to the optimisation problem in \eqref{eq:optimization}.

If $f_k$ is strictly convex, so is the optimisation problem, whose unique solution can therefore be obtained by applying the Karush–Kuhn–Tucker conditions \cite{tabak1971optimal} to the Lagrangian function
$\mathcal{L}=\sum_{k=1}^m f_k(i_k) - \lambda^\top (Bi - \bar{d})$,
with Lagrange multipliers $\lambda \in \mathbb{R}^n$. The cost functional $J(i)$ is thus directly linked to a physical Lagrangian, i.e., an energy function that accounts both for the energetic effort of the organism and for the constraints imposed by the boundary conditions, each weighted by one of the Lagrange multipliers (to be estimated). Requiring that $\mathcal{L}$ has zero derivative with respect to $i$, to identify the gradient along which to perform the minimisation, yields
$\nabla f(i) - \lambda^\top B = 0$.
The first derivative of the components of $f$ is $f_k' = g_k$, which is invertible with inverse $M_k$. Hence, we get $i_k = M_k(B^\top_k \lambda)$ and
\begin{equation}
    i = M(B^\top \lambda).
\end{equation}
The solution to the optimisation problem ~\eqref{eq:optimization} is thus the unique steady state of system~\eqref{eq:model_final}, satisfying the equilibrium condition $BM(B^\top \lambda) - \bar{d} = 0$, with the Lagrange multiplier being the asymptotic (steady-state) potential, $\lambda=\bar v= v(\infty)$ \cite{Blanchini2021}, and the asymptotic flow solves the optimisation problem in~\eqref{eq:optimization}. 

Hence, if $M_k$ are strictly increasing, and thus $f_k$ are strictly convex, the asymptotic endoplasm flow is optimal \textit{and} unique: the optimisation problem has a single solution.

On the other hand, if $M_k$ are non-decreasing, strict convexity is replaced by convexity, and hence, although the optimality result still holds, the optimal flow may not be unique. We further discuss this observation in Sec.~\ref{sec:branches}, as this property may lead to the appearance of multiple coexisting alternative paths. \\

Now, denote as $\mathbb{P}$ the set of all possible paths connecting the food sources. Is the input flow $d$ (endoplasm stemming from the main body located at the entrance) eventually channelled along the \textit{shortest path} $\mathcal{P}^* \in \mathbb{P}$? We can again resort to the circuit formalism in \cite{Blanchini2021} to analytically prove that this is indeed the case. If we consider the ideal threshold definition for $M_k$, with inverse given by \eqref{eq:M_k}, the functional in \eqref{eq:functional} becomes
\begin{equation}
    J^{th}(i) \doteq \sum_{k=1}^m V_{T_k} |i_k|,
    \label{eq:new_functional}
\end{equation}
and the optimisation problem in~\eqref{eq:optimization} becomes 
    \begin{align}
    \begin{split}
            \min \quad & \sum_{k=1}^m V_{T_k} |i_k|  \\
            \text{subject to } \quad & Bi = \bar{d}.
        \end{split}
        \label{eq:optimization2}
    \end{align}
Hence, the flows over the links are distributed so as to minimise the overall consumed power, measured as the product between the flow of endoplasm through a link and the threshold potential to initiate such flow.

To proceed, consider a positive input flow $d$ (associated with \textit{Physarum} entering the environment) in $\bar{d} = [d, 0, \dots, 0]^\top$ and, without loss of generality (since link orientation is arbitrary and embedded in $B$), assume that all the links in the network are oriented so that $i_k \geq 0$. Then, consider the modified optimisation problem
    \begin{align}
    \begin{split}
        \min \quad & \sum_{k=1}^m V_{T_k} i_k \\
        \text{subject to } \quad & Bi = \bar{d}, \\
                                & i \geq 0.
    \end{split}
        \label{eq:optimization3}
    \end{align}
The solution $i^*$ to problem~\eqref{eq:optimization2} is also
\begin{itemize}
\item a feasible solution to problem~\eqref{eq:optimization3}, since all the entries of $i^*$ are non-negative by construction;
\item an optimal solution to problem~\eqref{eq:optimization3}. In fact, assume by contradiction that a feasible solution $i^{**}$ exists that yields a smaller cost value for problem \eqref{eq:optimization3}; then, $i^{**}$ would be a feasible solution of \eqref{eq:optimization2} with a smaller cost, and hence it would be the actual optimal solution to \eqref{eq:optimization2}.
\end{itemize}
The solution to problem~\eqref{eq:optimization3} with a generic $d>0$ is a rescaling of the solution obtained when $d=1$, which corresponds to the \textit{shortest} path yielding the optimal cost $d \sum_{k \in \mathcal{P}^*} V_{T_k}$, as proven in \cite{magnanti1993network}. This solution ensures that the whole endoplasm flow is channelled through the minimum-cost path, which corresponds to the most energy efficient foraging behaviour for \textit{Physarum}.

Consequently, given the input flow $d$, when shuttle-streaming follows threshold mechanisms for decision-making, the steady-state flow, which solves the optimisation problem~\eqref{eq:optimization2}, corresponds to the whole flow connecting the food sources being directed along a path $\mathcal{P}^* \in \mathbb{P}$ that minimises the cost
\begin{equation}
J^{path}(\mathcal{P}) \doteq d \sum_{k \in \mathcal{P}} V_{T_k}, \quad \mathcal{P} \in \mathbb{P},
\end{equation}
which is the sum of all the thresholds in pressure potential, \textit{i.e.}, the total power required to form and maintain new branches, associated with the links along the chosen path through which endoplasm flows; this is precisely what happens in Fig.~\ref{fig:sims}a,b, which can now be fully understood in terms of optimisation with respect to an energetic cost functional.

Even when the cost functional is not strictly convex, and hence uniqueness of the solution is not guaranteed, the resulting minimum path is unique with high probability if the parameters governing the network model are randomly selected (as it occurs in Fig.~\ref{fig:sims}b). In fact, 
in the presence of randomly distributed heterogeneity in the environment,
the probability of finding more paths with the same total cost is very low, and thus a unique path asymptotically emerges with high probability. On the other hand, whenever pressure or chemical gradients lead to structured environments with special symmetries, more paths may happen to minimise the cost function, yielding the emergence of multiple branches at steady state. Conversely, in homogeneous environments, the minimum path is a straight line as in Fig.~\ref{fig:sims}a. \\

So far, we have shown that the minimum path is obtained when considering \textit{ideal} threshold functions as characteristic functions.
Nonlinear responses are necessary for minimum path formation, as proven in Sec.~\ref{sec:linear}.
Then, Sec.~\ref{sec:branches} shows that, if we consider approximations of the threshold function, such as those in Fig.~\ref{fig:characteristic_func}, the closer the characteristic functions are to the ideal threshold, the closer the flow is to being entirely conveyed along the minimum-path; the discussion in Sec.~\ref{sec:branches} also explains the formation of secondary branches that can be observed in reality (and also in Fig.~\ref{fig:sims}c), when the functions are not ideal thresholds \cite{Ghanbari2023}.

\subsection{Linear responses}\label{sec:linear}
To prove the necessity of nonlinear mechanisms to ensure optimality, we assess what happens if the characteristics of all memristors are linear, namely, if we have classical linear resistors with resistance $R_k$ associated with the links, so that $M_k(v_k)= v_k/R_k$. This is the functional form considered in scenario (d) of Sec.~\ref{sec:numerics}, where $R_k=6.7\cdot10^{-3}$. This case would correspond to \textit{Physarum} relying on simple memory-less mechanisms to form and maintain branches, and diffusing uniformly in the whole environment -- without the feedback reinforcement mechanism, brought about by flows of nutrients, that has been observed in experiments \cite{Tero2007}. Recalling the biophysical transition from gel to sol governed by pressure gradients discussed in Sec.~\ref{sec:2}, assessing whether a nonlinear (threshold) or a linear characteristic function is needed to capture this phenomenon amounts to asking: does the pressure gradient increase up to the point when it suddenly causes the gel to break down into sol to form new low-viscosity channels, or can channel formation be induced gradually also by lower pressure gradients?

Within the considered modelling framework, we can formally prove that the formation of optimal paths requires characteristic functions that are not linear. In fact, with linear characteristic functions $M_k(v_k)= v_k/R_k$, we have $g_k(i_k)=M_k^{-1}(i_k)=R_k i_k$ and the cost functional in \eqref{eq:functional} becomes
\begin{equation}
    J(i) = \sum_{k=1}^m \int_0^{i_k} R_k I dI = \frac{1}{2} \sum_{k=1}^m R_k i_k^2.
\end{equation}
The solution to the corresponding optimisation problem is well known in the literature on networks of resistors: the optimal solution minimises the total dissipated power by uniformly distributing currents all over the network, as shown e.g. in \cite[Application 1.8]{magnanti1993network}, which is the opposite of concentrating currents along a single path. The biological interpretation is the following: due to the linearity of the characteristic functions, the spatial distribution of endoplasm aimed at minimising the total dissipated power leads to flows that, instead of forming well-defined paths that solve the maze, diffuse all over the network corresponding to the maze environment. This is precisely what is shown in Fig.~\ref{fig:sims}d, where the absence of the nonlinear mechanism makes the organism diffuse uniformly in the environment.
Nonlinear responses are thus necessary to explain \textit{Physarum}'s ability to solve mazes and form minimum paths.

\subsection{Secondary branches}
\label{sec:branches}

Despite its impressive maze-solving and minimum-path-formation abilities, in reality \textit{P. polycephalum} may display some secondary branches that are still exploring the environment, or connecting food sources through alternative non-minimum paths, even after long experimental times \cite{Jones2010,Nakagaki2007,Ghanbari2023, reginato2025bottom}. 
The existence of these secondary branches can have a twofold explanation in light of the considered dynamical model. On the one hand, as discussed above, the optimal flow is not unique if $M_k$ are non-decreasing (instead of being strictly increasing) and hence the functional is convex, but not strictly convex, as in \eqref{eq:new_functional}. On the other hand, the characteristic functions $M_k$ are unlikely to be ideal thresholds, and more likely to be close approximations, such as those in Fig.~\ref{fig:characteristic_func}; deviations from ideal threshold functions may thus yield slightly deviating branches with respect to the minimum-path flow.

Still, provided that the steady-state flows corresponding to the characteristic functions are uniquely defined, the closer the characteristic functions are to ideal thresholds, the closer the steady-state flows are to the minimum-path flow. In particular, given a sequence of monotonically increasing characteristic functions $\{M_k^{(j)}\}_{j \in \mathbb{N}}$ that converge to the ideal threshold function $M_k^{th}$ as $j \to \infty$ (consider e.g. functions of the form \eqref{eq:char_func2} with increasing $j \in \mathbb{N}$) and are associated for all $j$ with uniquely defined steady-state flows in the links, the sequence of steady-state flows $\bar{i}_k^{(j)}$ (associated with $M_k^{(j)}$) converges to the optimal flow ${i}_k^*$ (associated with $M_k^{th}$) that flows along the minimum path, provided that such path is unique. Formally, if $\lim_{j \to \infty} \|M_k^{(j)} - M_k^{th}\| = 0$, and if the minimum path is unique, then 
\begin{equation}
    \lim_{j \to \infty} \|\bar{i}_k^{(j)} - {i}_k^*\| = 0.
\end{equation}
The steady-state flow $\bar{i}_{k}^{(j)}$ with characteristic function $M_k^{(j)}$ corresponds to the minimiser of problem~\eqref{eq:optimization}, with cost $\sum_{k=1}^m f_k^{(j)}(i_k) = \sum_{k=1}^m \int_0^{i_k} [M_k^{(j)}]^{-1}(I)dI$.
The ideal steady-state flow ${i}_{k}^*$ with characteristic function $M_k^{th}$, such that all entering endoplasm $d$ flows along the minimum path identified in Sec.~\ref{sec:opt_path}, corresponds to the minimiser of problem~\eqref{eq:optimization2}. The proof, provided in Appendix~\ref{appendix:A}, follows the same reasoning as in \cite[Proof of Theorem~1]{Blanchini2021}, by exploiting the circuital analogy.

This result guarantees that, if the decision-making mechanisms are close enough to an ideal threshold response, then the flow of endoplasm tends to form the minimum-path connection, regardless of the specific characteristic functions. However, as \textit{Physarum}'s dynamics is close but not perfectly identical to that of memristor models \cite{braund2016physarum}, such small deviations from the ideal case can explain the small alternative branches sometimes observed in experiments \cite{Jones2010}. The case of a non-ideal threshold function yielding secondary branches is also illustrated in Fig.~\ref{fig:sims}c. Non-ideal threshold responses may be beneficial to promote the formation of alternative branches in homogeneous environments \cite{Nakagaki2004}, in turn enabling robustness and flexibility through redundancy \cite{nakagaki2004smart, adamatzky2013creativity}. This observation suggests the hypothesis that a simple organism like \textit{Physarum} can pursue path optimisation and at the same time leave some room for adaptation though redundancy, without requiring two specific mechanisms for the two goals, but simply relying on quasi-ideal threshold responses. Testing this hypothesis, suggested by dynamics instead of bio-mechanics, would require the design of targeted experiments.

\section{Conclusion}
\textit{Physarum polychephalum} has long attracted a wide interest across biology, soft-matter physics and cognitive science for its remarkable problem-solving abilities \cite{ntinas2017oscillation, Nakagaki2004, Tero2007}; it has been long studied and modelled as a paradigmatic organism for decentralised problem solving and distributed intelligence. In this work, we have adopted the perspective of dynamical systems on networks to show that \textit{Physarum}'s maze solving behaviours are analogous to those obtained by solving an optimisation problem on the corresponding network, provided that sensing and branch formation are governed by threshold mechanisms.

To focus on \textit{Physarum}'s self-organising abilities associated with minimum-path formation, we have employed a phenomenological model of its network dynamics; our results suggest that the considered network optimisation framework is effective at capturing \textit{Physarum}'s adaptive behaviour. We have identified the cost function to be minimised to reproduce \textit{Physarum}'s foraging behaviour, which considers the geometrical properties of the environment (e.g., a maze) and also balances the benefit of transporting nutrients with the cost of maintaining the endoplasm network.

Our approach considers \textit{Physarum} as a model organism and merges several different concepts and viewpoints (dynamics on graphs and circuital analogies, optimisation theory and biological interpretation) to address a long-standing hypothesis in systems biology: intelligent problem solving by simple organisms can be achieved as the emergent solution to the decentralised optimisation of nonlinear dynamics, without requiring a specialised central nervous system. Although this hypothesis has been investigated in the literature \cite{Beekman2015,Gao2019,Adamatzky2016}, a formal mathematical proof in its support, based on a dynamical model and on the identification of a functional to be optimised, was previously missing to best of our knowledge. Our mathematical results are aligned with the experimental evidence showing that \textit{Physarum} performs shuttle-streaming across food sources and mazes of different topologies by following optimal paths, without requiring explicit computation or sensory awareness, but through diffusion in structured environments followed by threshold-driven refinement \cite{adamatzky2013creativity, Nakagaki2000, adamatzky2012slime}.

We have also identified the fundamental dynamical mechanism needed to mimic \textit{Physarum}'s problem-solving abilities, a result that is particularly useful in the synthetic design of bio-inspired circuits.
In particular, we have shown that threshold mechanisms for sensing and branch formation are necessary to achieve optimal paths, whereas responses that are too different from threshold functions fail at steering the flow to a single, optimal path, which explains the appearance of secondary branches.
We have also shown that getting ``sufficiently close" to ideal thresholds still allows to solve mazes and connect food sources: this is important in the context of synthetic \textit{Physarum}-inspired circuits (e.g., of memristor networks \cite{Pershin2011, ntinas2018coupled} or of bio-replicas \cite{ntinas2017oscillation, adamatzky2012slime, guo2020pora}), because it confirms that the desired problem-solving abilities can be obtained by realistic devices, where threshold functions are not ideal.

The main limitations of our work stem from the use of a phenomenological memristor-based model of the dynamics of \textit{P. polycephalum}, which prevents us from capturing other aspects that are still crucial for its development and survival. For instance, the threshold mechanisms lump together shuttle-streaming towards food or away from stressors such as light sources \cite{Nakagaki2007}, which cannot be distinguished in our model. In addition, we neglect non-equilibrium effects, such as subsequent path reinforcement via addition of nutrients or exogenous stress. Our dynamical model of \textit{Physarum} on a network also neglects the initial exploration phase in which the mould partially covers the environment, which is assumed to be already concluded, and focuses on the subsequent refinement of exploratory and linking branches with specified boundary conditions and static topologies; extending the analysis to encompass all the different phases of \textit{Physarum}'s evolution would require alternative modelling approaches \cite{liu2017new, reginato2025bottom, Ghanbari2023} that would not be amenable for formal analysis, and is left for future studies.

Another limitation stemming from the phenomenological nature of the model is that, albeit it captures key dynamical properties of \textit{Physarum}'s behaviour, its parameters do not exactly correspond to measurable biological quantities, which challenges the verification of the model predictions in a biological context. In particular, concerning the meaning of the characteristic function $M_k$ for the physiology of \textit{Physarum}, past studies suggest it to be linked to gel breaking down into sol after hitting thresholds of pressure gradients \cite{matsumoto2008locomotive} or the nonlinear interplay between the mechanics of the cytoplasm and flow velocity \cite{Ghanbari2023}, or combinations thereof; however, future studies are needed to precisely associate $M_k$ with biological properties, which may be investigated by empirically measuring the response of \textit{Physarum}'s branches to pressure stress, as recently done for instance in \cite{schmidt2025electrical}.

Finally, our approach does not provide quantitative predictions of \textit{Physarum}'s development, since multiple mechanisms may further concur to guide the initial exploration phase and its subsequent refinement, and to shape the characteristic function $M_k$. Our results are only valid qualitatively, to confirm the emergence of global optimisation abilities that, after an initial exploration phase, lead \textit{Physarum} to stream along a single optimal path, despite alternative routes being potentially available. Although our qualitative results hold irrespective of the exact functional form of $M_k$, finding a precise quantitative characterisation and connection to \textit{Physarum}'s mechanochemical properties would be fundamental to enable predictions of its evolution; as mentioned, however, this endeavour is demanded to future mechanistic and empirical studies.

Overall, our work supports the idea that maze solving by primitive organisms can be understood in terms of spontaneous global optimisation \cite{boettcher2000nature}, emerging from local decision-making brought about by collective nonlinear responses over networks. Our observations about collective intelligence can inspire the development of problem-solving algorithms that leverage the circuital model we have employed; our results could be extended to artificial systems emulating biological circuits via analog-digital technologies \cite{Kolka2019}, or to synthetic networks of memristors tuned to solve problems ranging from statistics to environment exploration \cite{Adamatzky2016}. They could also translate into bio-inspired algorithms for applications including task optimisation \cite{sun2017physarum} and logistics \cite{chu2021physarum}.

\vspace{3pt}

\section*{Acknowledgements} 
This work was funded by the European Union through the ERC INSPIRE grant (project number 101076926). Views and opinions expressed are however those of the authors only and do not necessarily reflect those of the European Union or the European Research Council. Neither the European Union nor the granting authority can be held responsible for them.

The authors would like to thank David Palma for the access to a previously optimised \cite{Blanchini2021} ODE solver. 

Conceived and designed the research: G.G.
Derived the model: D.P.
Performed numerical simulations: D.P.
Derived mathematical results: G.G.
Interpreted the results: D.P., G.G. 
Supervised the project: G.G. 
Wrote the paper: D.P., G.G.

The authors declare no competing interests.

\section*{Data Availability} 
The code for the numerical simulations that support the findings of this article is available upon request.

\appendix
\section{Proof of the result in Sec.~\ref{sec:branches}} 
\label{appendix:A}
We provide, for completeness, the proof of the result stated in Sec.~\ref{sec:branches}: given a sequence $\{M_k^{(j)}\}_{j \in \mathbb{N}}$ of characteristic functions that approach the ideal threshold function $M_k^{th}$ as $j \to \infty$, the steady-state flow $\bar{i}_k^{(j)}$ (associated with $M_k^{(j)}$) approaches the optimal flow ${i}_k^*$ (associated with $M_k^{th}$) along the minimum path, under the assumption that such a minimum path is unique.
We leverage the circuital interpretation so as to follow the reasoning proposed for electrical currents in \cite{Blanchini2021, magnanti1993network}, which share the same mathematical framework.

The steady-state flow $\bar{i}^{(j)} = [\bar{i}_1^{(j)} \dots \bar{i}_m^{(j)}]^\top \in \mathbb{R}^m$ obtained for $M^{(j)}(\cdot)= [M_1^{(j)}(\cdot) \dots M_m^{(j)}(\cdot)]^\top \in \mathbb{R}^m$ is defined as the unique solution to the optimisation problem 
\begin{align}
\begin{split}
    \min \quad & \sum_{k=1}^m f_k^{(j)}(i_k)\\
            \text{subject to } \quad & Bi = \bar{d},
\end{split}
\label{eq:optimization_j}
\end{align}
with cost functional
\begin{equation}
    J^{(j)}(i) = \sum_{k=1}^m f_k^{(j)}(i_k),
\end{equation}
where $f_k^{(j)}(i_k) = \int_0^{i_k} [M_k^{(j)}]^{-1}(I)dI$.
The minimiser $\bar{i}^{(j)}$ is unique because functions $M_k^{(j)}$ are assumed to be strictly increasing for all $k$, and hence $f_k^{(j)}$ and $J^{(j)}$ are strictly convex.

The ideal optimal flow ${i}^* = [{i}_1^* \dots {i}_m^*]^\top \in \mathbb{R}^m$ is instead the solution to the optimisation problem \eqref{eq:optimization2}, with cost functional $J^{th}(i)=\sum_{k=1}^m V_{T_k} |i_k|$ defined in \eqref{eq:new_functional}. We denote by $\bar{J}^{th} = J^{th}({i}^*)$ the optimal value of the cost corresponding to the optimal flow.

The proof is conducted in two steps: (I) show that all optimal solutions are uniformly bounded; (II) show that such solutions actually converge to the ideal one.

(I) For any given $z$, $\lim_{j \to \infty} \|f_k^{(j)}(z) - V_{T_k} |z|\|=0$. Then, $\lim_{j \to \infty} \| J^{(j)}(z) - J^{th}(z)\|=0$ and, in particular, $\lim_{j \to \infty} \| J^{(j)}(i^*) - J^{th}(i^*)\|= \lim_{j \to \infty} \| J^{(j)}(i^*) - \bar{J}^{th}\| =0$.
Hence, the sequence $\{J^{(j)}(\bar{i}^{(j)}) \}_{j \in \mathbb{N}}$ of optimal costs is upper bounded by the sequence $\{J^{(j)}(i^*) \}_{j \in \mathbb{N}}$ that converges to $\bar{J}^{th}$ as $j \to \infty$, namely,
\begin{equation}
    J^{(j)}(\bar{i}^{(j)}) \leq J^{(j)}(i^*) \xrightarrow[j \to \infty]{} \bar{J}^{th}.
    \label{eq:proof1}
\end{equation}
All cost functionals ($J^{(j)}$, for all $j$, and $J^{th}$) are radially unbounded, since they are sums of non-negative radially unbounded functions ($f_k^{(j)}$ and $V_{T_k} |i_k|$). Hence, the corresponding optimal solutions $\bar{i}^{(j)}$ and $i^*$ are finite.
By \eqref{eq:proof1}, there exists a value $\kappa > 0$ such that $\bar{i}^{(j)}$ are inside the compact set
${\mathcal{S}}^{(j)} = \{u \in \mathbb{R}^m | {J}^{(j)}(u) \leq \kappa \}$.
Since the sequence of cost functionals $J^{(j)}$ converges to the radially unbounded $J^{th}$, the sequence of sets ${\mathcal{S}}^{(j)}$ is uniformly bounded in a compact set $\mathcal{S}$. Therefore, $\{\bar{i}^{(j)} \}_{j \in \mathbb{N}} \in \mathcal{S}$, meaning that all optimal solutions are uniformly bounded.

(II) To prove convergence of the flows $\bar{i}^{(j)}$ to the ideal flow $i^*$, $\lim_{j \to \infty} \| \bar{i}^{(j)} - i^* \|=0$, assume by contradiction that $\lim_{j \to \infty} \| \bar{i}^{(j)} - i^* \| \neq 0$. This implies the existence of an open neighbourhood $\mathcal{U} \subset \mathcal{S}$ of $i^*$ and of a sub-sequence of $\{ \bar{i}^{(j)} \}_{j \in \mathbb{N}}$ confined in the compact set $\mathcal{S} \setminus \mathcal{U}$, which must admit a sub-sub-sequence converging to some $i^\circ \in \mathcal{S} \setminus \mathcal{U}$, with $i^\circ \neq i^*$.
All elements $\bar{i}^{(l)}$ of the sub-sub-sequence are solutions to \eqref{eq:optimization_j} and thus satisfy the constraint $B \bar{i}^{(l)} = \bar{d}$; hence, also the limit $i^\circ$ satisfies $Bi^\circ = \bar{d}$.
To obtain a contradiction, we now show that 
\begin{equation}
    J^{th}(i^\circ) \leq \bar{J}^{th},
    \label{eq:contr}
\end{equation}
which implies either that $\bar{J}^{th}$ is not the optimal value of the cost (if $J^{th}(i^\circ)$ is strictly smaller), thus contradicting the optimality assumption, or that the optimisation problem \eqref{eq:optimization2} has two minima (both $i^*$ and $i^\circ$), thus contradicting the uniqueness assumption.

To obtain \eqref{eq:contr}, observe that $\lim_{j \to \infty} g_k^{(j)}(i_k) = V_{T_k}$. Then, being the set $\mathcal{S}$ bounded, there exists a finite $\gamma>0$ such that $g_k^{(j)}(i_k) \leq \gamma$, for all $i_k$ such that $i \in \mathcal{S}$, for all $k$ and for all large enough $j$.
Hence, when $j$ is large enough, the gradient of $J^{(j)}$ is uniformly bounded: for any $z$,
\begin{equation}\label{eq:boundedgradient}
    \left| \frac{\partial J^{(j)}}{\partial i_k} \right|(z) = g_k^{(j)}(z) \leq \max_{i_k \text{ s.t. } i \in \mathcal{S}} \{g_k^{(j)} (i_k) \} \leq \gamma.
\end{equation}
Since $J^{(j)}$ is convex for all $j$, \eqref{eq:boundedgradient} implies the existence of a constant $L$ such that, for large enough $j$,
\begin{equation}
    |J^{(j)}(u_1) - J^{(j)}(u_2)| \leq L \|u_1 - u_2\|, \quad \forall u_1,u_2 \in \mathcal{S}.
    \label{eq:const}
\end{equation}
Then, for $\ell \in \mathbb{N}$ large enough, we have
\begin{equation}\label{eq:convergence}
\begin{array}{l}
    J^{th}(i^\circ) \\
    = J^{th}(i^\circ) - J^{(\ell)}(i^\circ) + J^{(\ell)}(i^\circ) - J^{(\ell)}(\bar{i}^{(\ell)}) + J^{(\ell)}(\bar{i}^{(\ell)}) \\ 
    \leq |J^{th}(i^\circ) - J^{(\ell)}(i^\circ)| + |J^{(\ell)}(i^\circ) - J^{(\ell)}(\bar{i}^{(\ell)})| + J^{(\ell)}(\bar{i}^{(\ell)}).
\end{array}
\end{equation}
As shown at step (I), $\lim_{\ell \to \infty} |J^{th}(i^\circ) - J^{(\ell)}(i^\circ)|=0$. In view of \eqref{eq:const}, $|J^{(\ell)}(i^\circ) - J^{(\ell)}(\bar{i}^{(\ell)})| \leq L \|i^\circ - \bar{i}^{(\ell)}\|$, and $\lim_{\ell \to \infty} \|i^\circ - \bar{i}^{(\ell)}\|=0$ by construction of the convergent sub-sub-sequence with limit $i^\circ$.
Hence, \eqref{eq:convergence} yields
\begin{equation}
    J^{th}(i^\circ) \leq J^{(\ell)}(\bar{i}^{(\ell)}) \leq J^{(\ell)}(i^*)  \xrightarrow[l \to \infty]{} \bar{J}^{th} 
\end{equation}
by \eqref{eq:proof1}, since $\{\bar{i}^{(\ell)} \}$ is a sub-sub-sequence of $\{\bar{i}^{(j)} \}$. This shows \eqref{eq:contr} and concludes our proof by contradiction.
Hence, flows converge to the ideal optimal flows along the minimum path, as the corresponding characteristic function converges to the ideal threshold function.

\bibliography{refs.bib}

\end{document}